\documentclass{ptephy_v1}%%%%%% to generate preprint number with ptep logo

\preprintnumber{XXXX-XXXX} %%% %%% Insert preprint number here

\begin{document}

\title{Air Shower Observation by a Simple Structured Fresnel lens
  Telescope with Single Pixel for the Next Generation of Ultra-High
  Energy Cosmic Ray Observatory}

%%%% To generate auto affiliation numbers please use \author{}\affil{} command

\author{Yuichiro Tameda}
\affil{Osaka Electro Communication University, Neyagawa, Osaka, Japan \email{tameda@osakac.ac.jp}}

\author{Takayuki Tomida}
\author{Mashu Yamamoto}
\author{Hirokazu Iwakura}
\affil{Shinshu University, Nagano, Nagano, Japan}

\author[3,4]{Daisuke Ikeda} %%% Use optional bracket [3] to change the respective address
\affil[3]{Institute for Cosmic Ray Research, The University of Tokyo, Kashiwa, Chiba, Japan}
\affil[4]{Earthquake Research Institute, University of Tokyo, Bunkyo-ku, Tokyo, Japan}

\author[5]{Katsuya Yamazaki}
\affil[5]{Kanagawa University, Yokohama, Kanagawa, Japan}

%%% To include the collaborator name... Please use the command "\collaborator"
%%% For example: \collaborator{ATLAS Collaboration}

\begin{abstract}%
  Improved statistics and mass-composition-sensitive observation
  are required to clarify the origin of ultra-high energy cosmic rays (UHECRs).
  Inevitably in the future,
  the UHECR observatories will have to be expanded due to the small flux;
  however, the upgrade will be expensive with the detectors currently in use.
  Hence, we are developing a new fluorescence detector for UHECR observation.
  The proposed fluorescence detector,
  called cosmic ray air fluorescence Fresnel-lens telescope (CRAFFT),
  has an extremely simple structure and can observe the longitudinal development of an air shower.
  Furthermore,
  CRAFFT has the potential to significantly reduce costs
  for the realization of a huge observatory for UHECR research.
  We deployed four CRAFFT detectors at the Telescope Array site and conducted the test observation.
  We have successfully observed ten air-shower events using CRAFFT.
  Thus, CRAFFT can be a solution to realize the next generation of UHECR observatories.
\end{abstract}

\subjectindex{F00, F03}

\maketitle

\section{Introduction}
Since the discovery of cosmic rays in 1912 by V.F.~Hess,
cosmic rays over a wide range of energies have been observed.
However, the origin of cosmic rays,
especially those with energies above $10^{18}~{\rm eV}$, has not yet been clarified.
In this regard, understanding the UHECR energy spectrum,
mass composition, and anisotropy is crucially important.
AGASA \cite{AGASASpectrum} reported that the energy spectrum of cosmic rays extends beyond $10^{20}~{\rm eV}$
without the Greisen-Zatsepin-Kuzmin (GZK) limit \cite{GZK}, 
even though the energy spectrum of HiRes \cite{HiResSpectrum} is consistent with the GZK cut-off.
In recent years,
the Pierre Auger Observatory (Auger) \cite{PAOSpectrum},
and Telescope Array (TA) \cite{TASpectrum} confirmed this energy spectrum,
demonstrating a flux suppression consistent with the GZK limit.
TA also reported that an intermediate-scale of anisotropy of cosmic rays with energies
greater than $57~{\rm EeV}$ exists in the northern sky, 
called as the hotspot \cite{TAHotspot}.
In addition,
Auger reported that
the direction of arrival of cosmic rays with energies above $8~{\rm EeV}$
has a dipole structure \cite{PAODipole},
indicating that the sources of such high-energy cosmic rays should be outside our galaxy.
Ultra-high energy cosmic rays (UHECRs) with energies above $10^{19}~{\rm eV}$
and a light nucleus (e.g. a proton)
can propagate with only a few degrees of deflection in the galactic magnetic field,
which is a few micro gauss.
As a result,
it is expected that UHECR sources can be identified.
Currently,
HiRes and TA have reported that the composition of UHECRs above $\sim{\rm EeV}$
is dominated by or consistent with light components such as proton \cite{HiResComp2010}\cite{TAComp2018}.
On the other hand,
Auger has indicated transition in the composition,
from a light to heavy component at energies above $10^{18.3}~{\rm eV}$ \cite{Auger2014}. 
However,
TA and Auger $X_{\rm max}$ data are in good agreement within the systematic uncertainties
at the moment \cite{WGComp2016}.
To understand UHECR composition and identify its sources,
higher statistics and composition-sensitive observation are indispensable.

In the future,
the expansion of UHECR observatories to obtain better statistics will be inevitable,
because the flux of UHECR with energies above $10^{20}~{\rm eV}$ 
is as low as a few events per $1000~{\rm km^2}$ per year.
To realize these inevitable upgrades with the detectors currently in use,
the cost will inflate in proportion to the effective area's extension.
Therefore,
cost reduction of detectors will be required.
Hence,
to help realize the required cost reduction,
we have developed a cosmic ray air fluorescence Fresnel lens telescope (CRAFFT)
based on the concept of a simple air fluorescence detector \cite{privitera}.
This concept is highly suitable for cost reduction to realize a huge ground array 
%to observe UHECRs
composed of fluorescence detectors
which can observe air shower longitudinal developments.
Other solutions,
such as a simple fluorescence detector,
to achieve a large effective area have also been proposed and are being studied.
For example,
the FAST project \cite{FAST} is developing telescopes comprising a
composite mirror and 8-inch Photomultiplier Tubes (PMTs).
Other candidates including fluorescence detectors
from space \cite{EUSO}\cite{TUS}\cite{POEMA} are also under development.
Herein,
we report the development and trial results of CRAFFT,
a novel fluorescence detector that is extremely simple and inexpensive
for the next generation of large-scale UHECR observatories.

\section{Cosmic Ray Air Fluorescence Fresnel lens Telescope (CRAFFT)}
The fluorescence technique adopted by CRAFFT was originally studied in the 1960s. 
At that time, various fluorescence detectors were developed, 
then the detection of the fluorescence light from air showers succeeded
by using Fresnel lens \cite{hara1970}.
After that,
not Fresnel lens but mirrors have been more adopted as light collectors for fluorescence detectors,
of which Fly’s Eye was one of the first successful example \cite{flyseye}.
Currently,
technological advancements have enabled easy achievement of large Fresnel lenses with high UV transmittance,
high-sensitivity photon sensors, and Flash ADC (FADC).
JEM-EUSO is one of the challenging project for UHECR observation
developing a high resolution Fresnel lens telescope
with large aperture and large field of view (F.O.V.) from space \cite{EUSO}. 
%Hence,
On the other hand, 
we are trying to develop a simple structure fluorescence detector
with a Fresnel lens for a huge ground array to observe UHECRs.

\subsection{Fluorescence Technique}
An air fluorescence detector measures scintillation light emitted from ${\rm N_2}$ molecules
in the atmosphere excited by energetic particles in extensive air showers that are induced by high-energy cosmic rays.
The wavelength band of the scintillation light ranges from $300$ to
$400~{\rm nm}$ with the peaks mainly at
$337.1~{\rm nm}$, $357.7~{\rm nm}$ and $391.4~{\rm nm}$.
Emitted light from an air shower is attenuated via Rayleigh and Mie scattering while passing through the atmosphere.
These faint UV photons are gathered and concentrated at the focal point of a fluorescence detector.

A modern fluorescence detector can observe an air shower track using
multi pixels such that TA fluorescence detector (FD) has 256 PMTs \cite{TAFD};
this allows the shower detector plane (SDP) to be determined.
The geometry of air showers can be determined by the arrival-timing information of each pixel.
When an air shower is stereoscopically observed by more than two fluorescence detectors at different sites,
its geometry can be determined as the line of SDP intersection.
In the case of a hybrid detector,
such as TA or Auger,
comprising fluorescence and surface detectors,
the additional information of arrival timing or shower core position provided by the surface detectors
can allow the geometry to be determined with an accuracy that is better than $1^{\circ}$.
Once the geometry of an air shower is determined,
the amount of light emitted onto the shower axis,
which is proportional to the energy loss of the shower particles,
can be calculated by considering the atmospheric attenuation.
Following this calculation,
the primary energy of cosmic rays can be estimated calorimetrically.

In order to determine the air shower geometry, 
a single-pixel Fresnel lens telescope such as CRAFFT detector
records time profile of fluorescence photons from the air shower using FADC together with at least two more detectors 
deployed separately with a spacing of $\sim 20~{\rm km}$. 
Without the information of scintillation detector or water Cherenkov detector array,
the geometry can be determined by arrival time information of photons.
Once the geometry is determined, the longitudinal development of energy deposit
as a function of atmospheric depth along the shower track can be reconstructed.
Then, the time profile of FADC considering the reconstructed longitudinal development can be simulated.
Using the simulated FADC time profile,
the shower geometry is fitted again to reproduce the recorded FADC time profile.
By the repetition of the above procedure,
the accuracy of the geometry and the longitudinal development will be improved
as a method of the profile constrained geometry fit technique adopted by the Fly's Eye experiment \cite{PCF}.
Then, the energy and the $X_{\rm max}$ can be determined.

The effective distance of a single-pixel Fresnel lens fluorescence detector
with $1~{\rm m^2}$ aperture and $16^{\circ} \times 16^{\circ}$ F.O.V. 
in which air showers of $10^{20}~{\rm eV}$ can be triggered is estimated to be up to $25~{\rm km}$ by the simulation \cite{tameda2016}. 
Therefore,
single-pixel FDs will be deployed as a huge array of a triangle lattice with $20~{\rm km}$ spacing \cite{privitera}.

\subsection{Detector configuration}

\begin{figure}[htbp]
  \centering
  \includegraphics[width=0.4\textwidth]{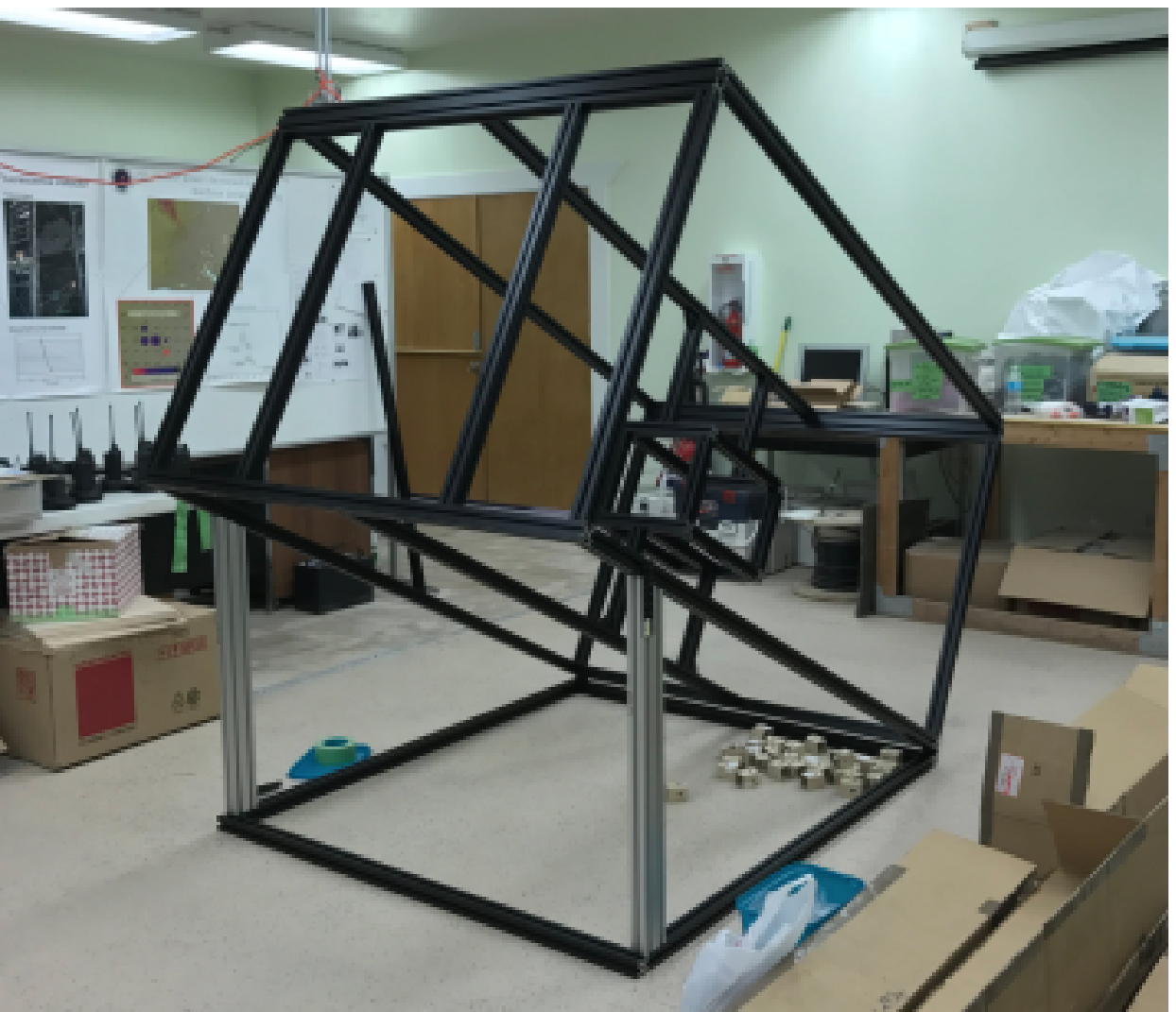}
  \hspace{0.5cm}
  \includegraphics[width=0.4\textwidth]{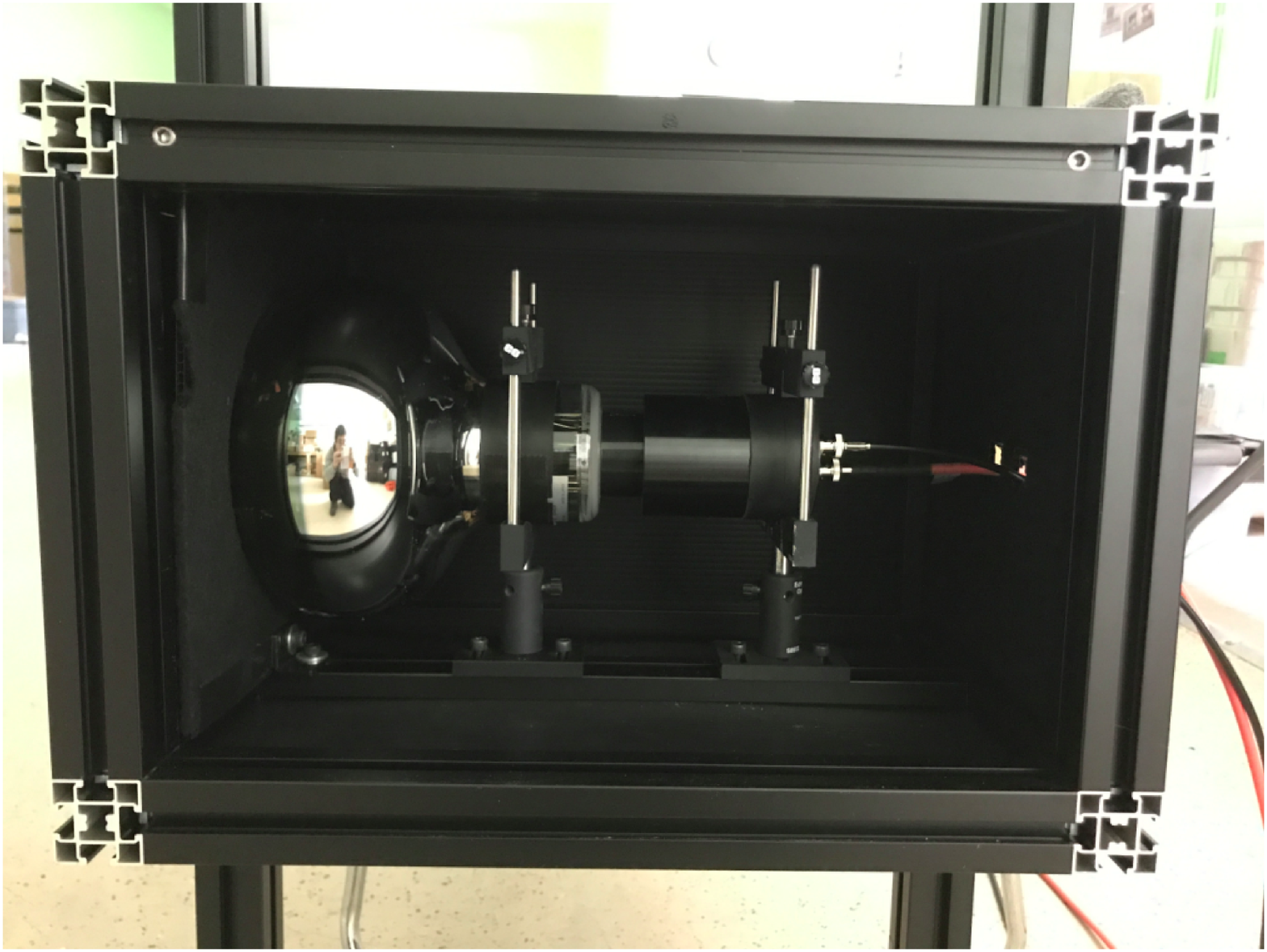}
  \caption{
    The structure of the CRAFFT detector made of anodized T-slotted aluminum extrusions.
    Left: At the aperture, two frames are attached to support the plastic
    lens to maintain the flatness.
    The elevation angle can be adjusted.
    Right: PMT mount installed at the focus of the Fresnel lens.
  }
  \label{fig-design}
\end{figure}

The structure of the CRAFFT detector is very simple,
which makes it cost-effective and easy to deploy.
Table \ref{component} lists the main components of the detector.
Figure \ref{fig-design} shows the structural design of the CRAFFT detector.
The CRAFFT detector's light collector is a Fresnel lens,
which is made of UV-transmitting acrylic plastic,
with a focal length of $1.2~{\rm m}$.
At the focal point,
an 8-inch PMT with a UV-transmitting filter is mounted.
The PMT signal is recorded by the 80-MHz sampling 12-bit
FADC board
with field-programmable gate array (FPGA).

\begin{table}[htb]
  \centering
  \caption{Component list of the CRAFFT detector}
  \label{component}
  \begin{tabular}{l|l}
    Component              & Product  \\ \hline \hline 
    Fresnel lens           & NTKJ, CF1200-B  \\ \hline 
    UV-transmitting filter & Hoya, U330  \\ \hline 
    Photomultiplier tube   & Hamamatsu, R5912 \\ \hline 
    HV power supply        & CAEN, N1470AR \\ \hline 
    HV divider             & (special order) \\ \hline 
    FADC board             & TokushuDenshiKairo, Cosmo-Z \\ \hline 
    Low pass filter        & Mini Circuit, BLP-15+ \\ \hline 
    Amplifier              & LeCroy, MODEL 612AM \\ \hline 
    Structure              & YUKI,  anodized T-slotted aluminum extrusions \\ \hline 
  \end{tabular}
\end{table}

\begin{figure}[htbp]
  \centering
  \includegraphics[height=0.34\textwidth]{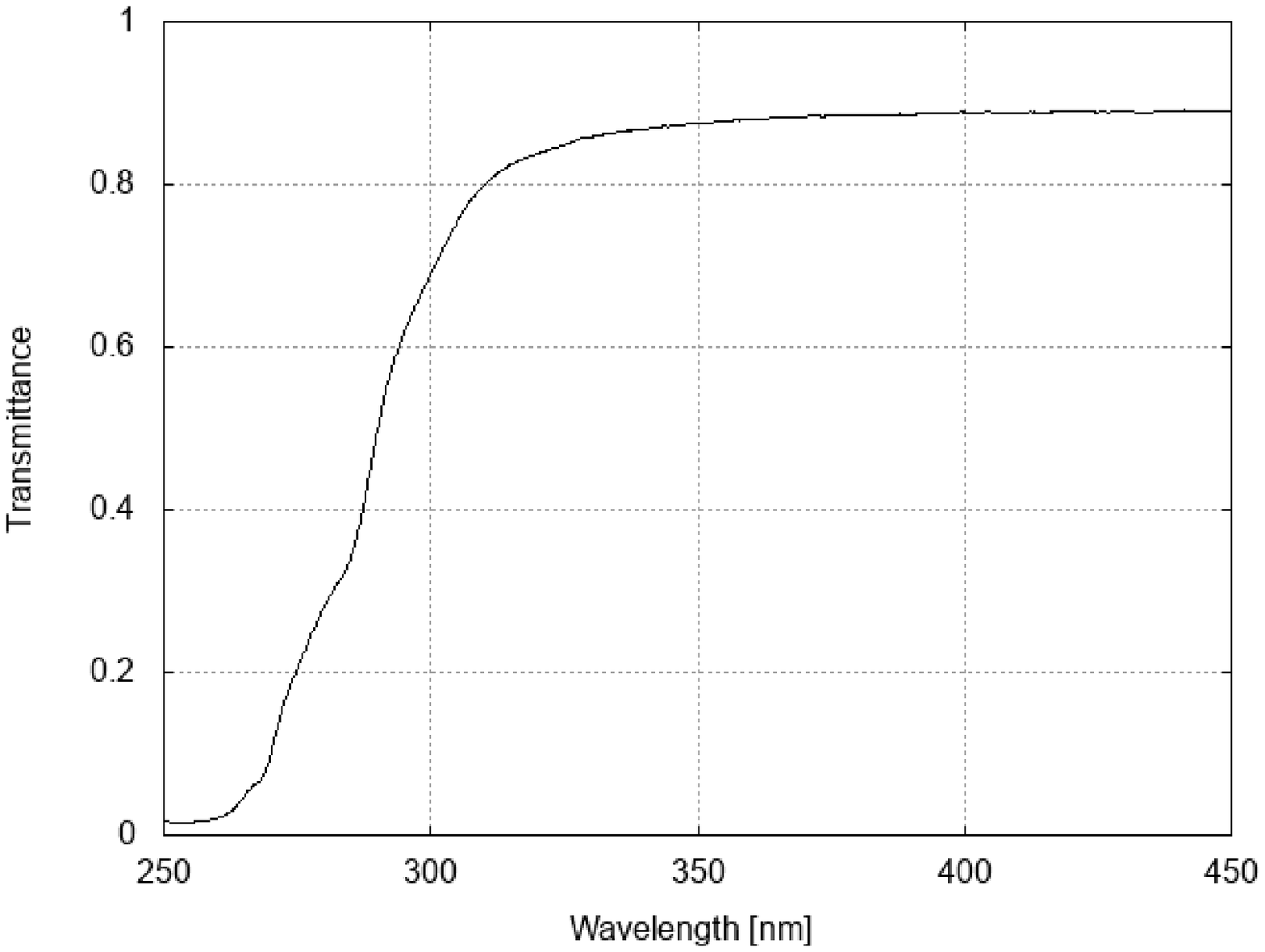}
  \includegraphics[height=0.34\textwidth]{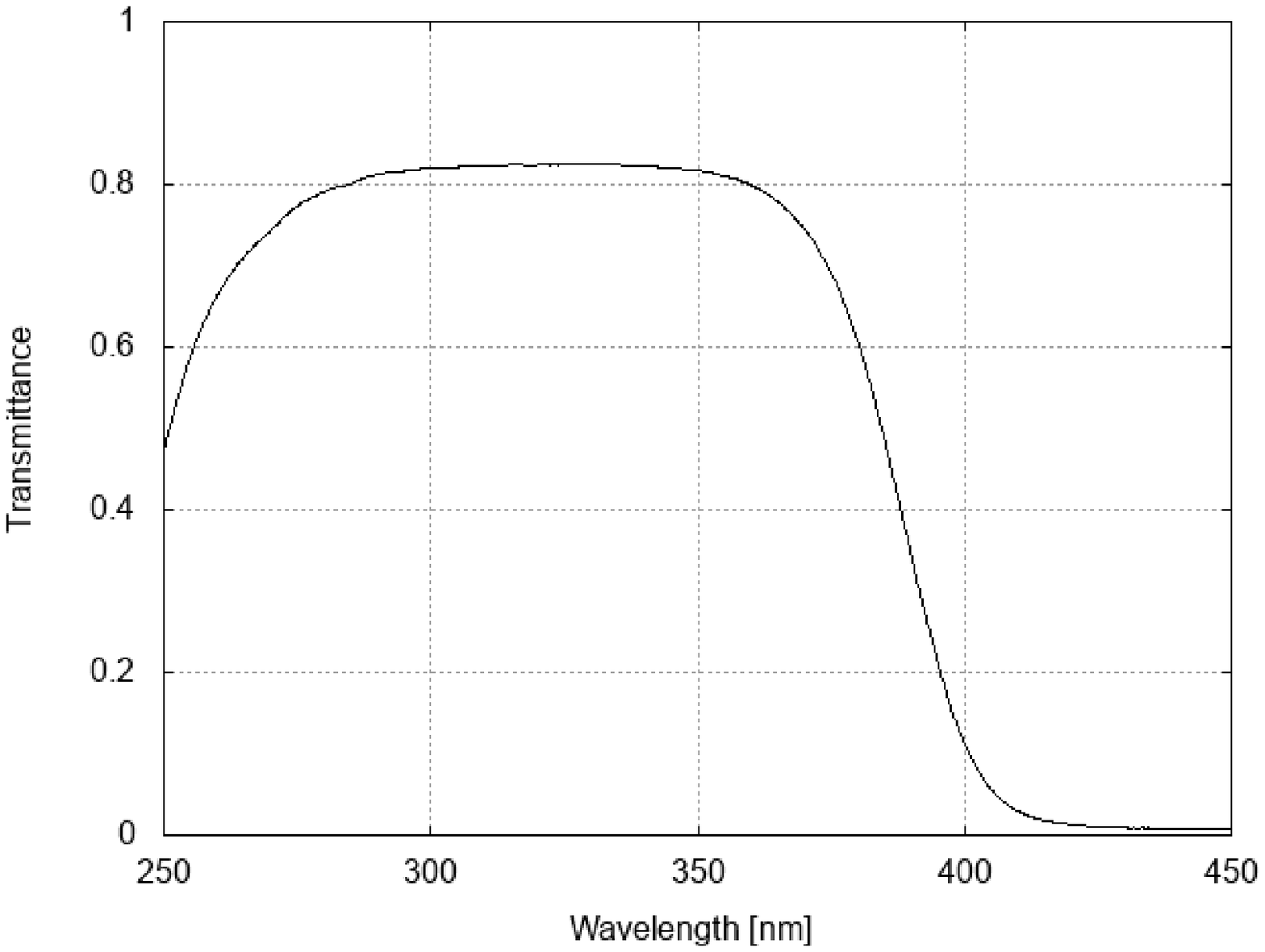}
  \caption{Transparency of PMMA which is the material of the Fresnel
    lens (Left) and 
    the UV-transmitting filter (Right) measured by
    a spectrometer (USB4000, Ocean Optics)
    with a mini deuterium halogen light source (DT-MINI-2-GS, Ocean Optics).}
  \label{fig-transparency}
\end{figure}

The focal length, effective aperture, dimensions, and pitch of the
Fresnel lens are
$1200~{\rm mm}$,
$\phi 1900~{\rm mm}$,
width $1400~{\rm mm}$, length $1050~{\rm mm}$, thickness $3~{\rm mm}$
and
$0.33~{\rm mm}$,
respectively.
The material of the Fresnel lens is acrylic plastic (Mitsubishi Chemical, ACRYLITE 000) and its transparencies
is shown in Fig. \ref{fig-transparency} on the left.
The scattering loss caused by the fine pitch of Fresnel lens is estimated 1.5\%
for the incident light with the angle of $4^{\circ}$ by ray tracing simulation.

To reduce night sky background light in the visible range that does not contribute to the signal,
a UV-transmitting filter of the range of $300 - 400~{\rm nm}$ is used.
The filter is flat, and its dimensions are
width $130~{\rm mm}$,
length $130~{\rm mm}$ and
thickness $2.5~{\rm mm}$.
The transparency is shown in Fig. \ref{fig-transparency} on the right.

\begin{figure}[htbp]
  \centering
  \includegraphics[width=0.3\textwidth]{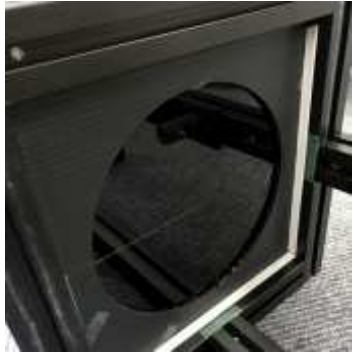}
  \caption{Spatial filter to limit the F.O.V. of the CRAFFT detector, 
    which in turn reduces the effect of the nonuniform photo detection efficiency
    due to the spherical surface of the PMT.}
  \label{fig-spatialFilter}
\end{figure}

The photon sensor of the CRAFFT detector is an 8-inch large photocathode area PMT with
a D type socket of a DC coupling with a negative high voltage power supply (E7694, Hamamatsu).
The spectral response and the peak wavelength of the PMT are
from $300~{\rm nm}$ to $650~{\rm nm}$ and $420 {\rm nm}$.
The photocathode and window material are bialkali and borosilicate glass, respectively.
The photon entrance window is a spherical surface; thus,
for this test observation,
we use a spatial filter of $16~{\rm cm}$ diameter to cut the incident
light at an angle of more than $\pm 4^{\circ}$ to reduce the ambiguity of
detection efficiency at the periphery of the photon entrance windows (Fig. \ref{fig-spatialFilter}).
The quantum efficiency at $390~{\rm nm}$ and the typical gain with $1500~{\rm V}$ of the PMT are 
$25~\%$ and $1.0 \times 10^{7}$, respectively.

We use a high-voltage power supply that is controlled via a network to
ensure that the Data acquisition (DAQ) system of CRAFFT can be controlled remotely.

The signal from air showers is so faint that 
we use amplifier to amplify the signal and 
low pass filter to reduce the high frequency night sky background.
The signal from the PMT is digitized and recorded by the FADC board
on which Linux is available with Zynq.
Zynq is a FPGA on which We can implement our own trigger algorithm now
we are developing.
The sampling rate and resolution of the Cosmo-Z are $80~{\rm MHz}$ and $12~{\rm bits}$, respectively.
We can time events with precise time stamps using a GPS module (Linx Technologies Inc. EVM-GPS-FM),
which provides 1 pulse per second as well as time information.

For the frame of the detector,
T-slotted aluminum extrusions,
that are black anodized to reduce light reflection, are used.
Aluminum extrusions are easy-to-build and modify;
hence,
this a reasonable choice for the prototype test.
To shield stray light and protect the components inside,
the detector is covered by an $0.4~{\rm mm}$ thick steel plate.
During daytime,
we pull down the roll curtain between the lens and the PMT to prevent incident light from the sun.

\begin{figure}[htbp]
  \centering
  \includegraphics[height=0.34\textwidth]{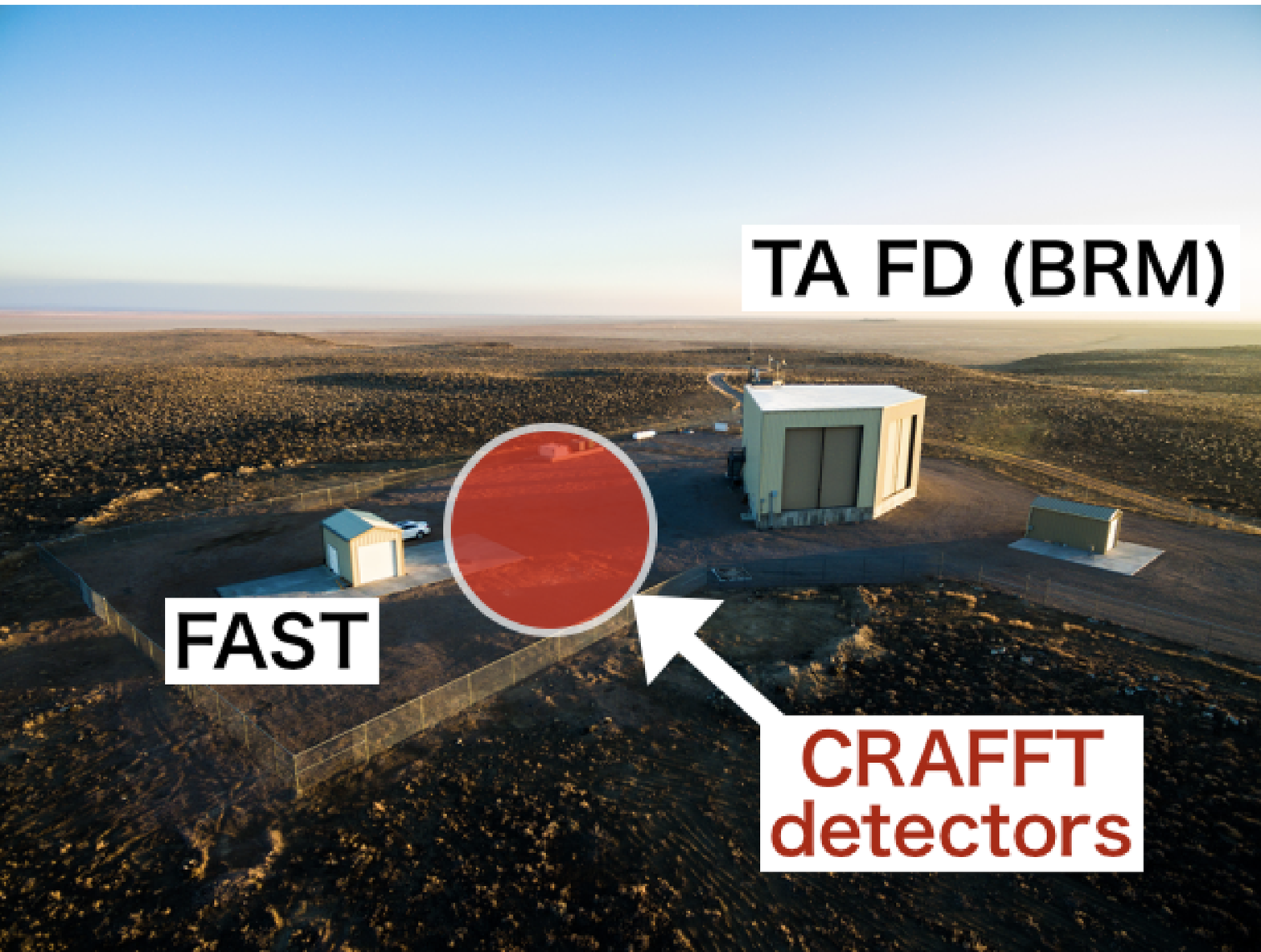}
  \includegraphics[height=0.34\textwidth]{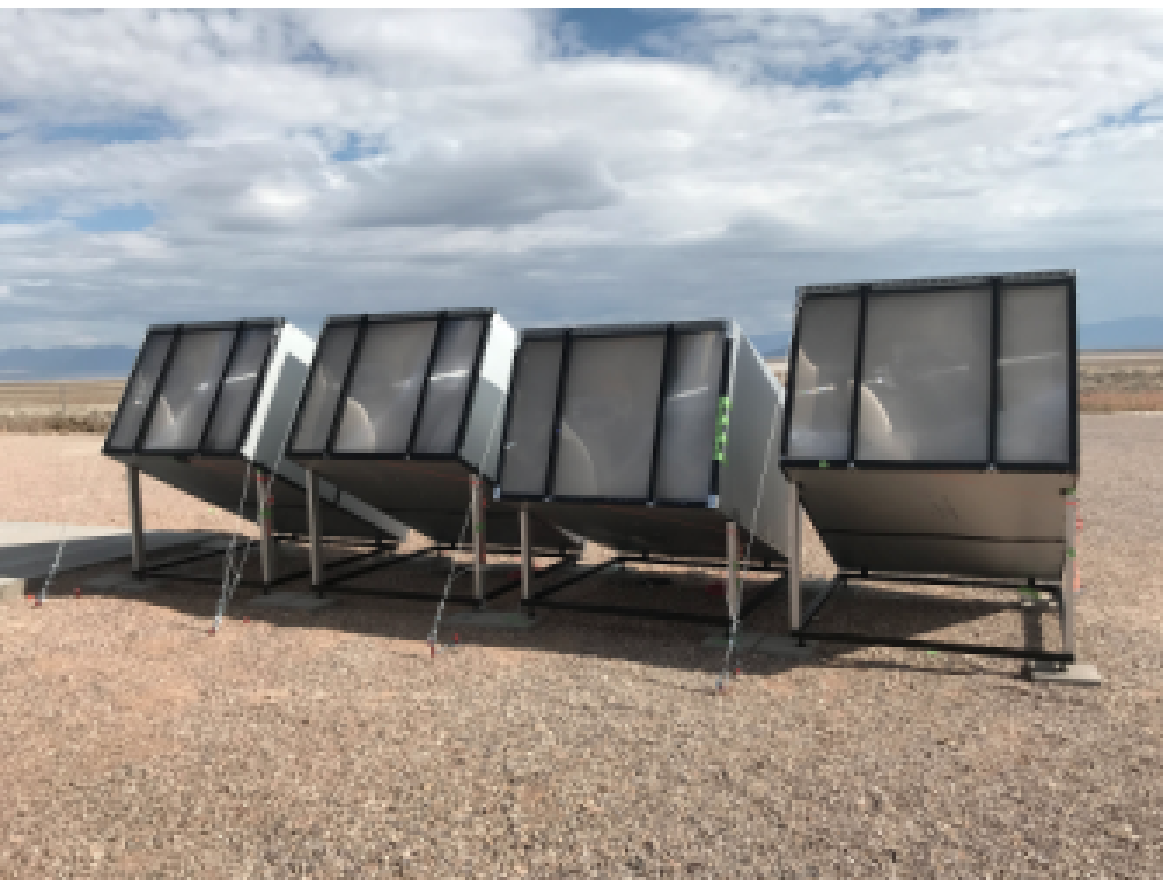}
  \caption{Left: 
    The location of CRAFFT next to the TA FD station.
    Right: Four deployed CRAFFT detectors.}
  \label{fig-deployed}
\end{figure}

\begin{figure}[htbp]
  \centering
  \includegraphics[width=0.95\textwidth]{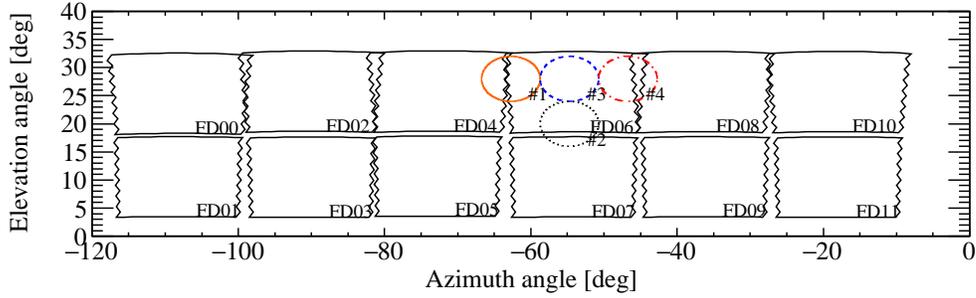}
  \caption{The F.O.V. of each CRAFFT detector is shown as circle drawn over
    the F.O.V. of the TA FDs edge of which is shown as a solid line.}
  \label{fig-FOV}
\end{figure}

In October 2017,
we deployed four CRAFFT detectors next to the fluorescence detector
station at the Black Rock Mesa site of the TA experiment in Utah U.S. (Fig. \ref{fig-deployed}).
The F.O.V. of each detector was $\pm 4^{\circ}$,
limited by the spatial filter (Fig. \ref{fig-spatialFilter}).
The elevation angles of the three detectors were $28^{\circ}$ and that of the other detector was $20^{\circ}$.
In this observation,
we optimized the arrangement of the detectors to observe relatively low-energy air showers
because the priority was to detect as many air shower events as possible.
The F.O.V. of the four detectors corresponded to the 6th fluorescence detector of TA, as shown in Fig \ref{fig-FOV}.
The upper center and lower detectors observe the vertical laser from the TA Central Laser Facility (CLF) \cite{tomida}.
The remaining upper-viewing two detectors observe at an angle $8^{\circ}$ east or west from the center detector.

\section{Observation}

\begin{figure}[htbp]
  \centering
  \includegraphics[width=0.9\textwidth]{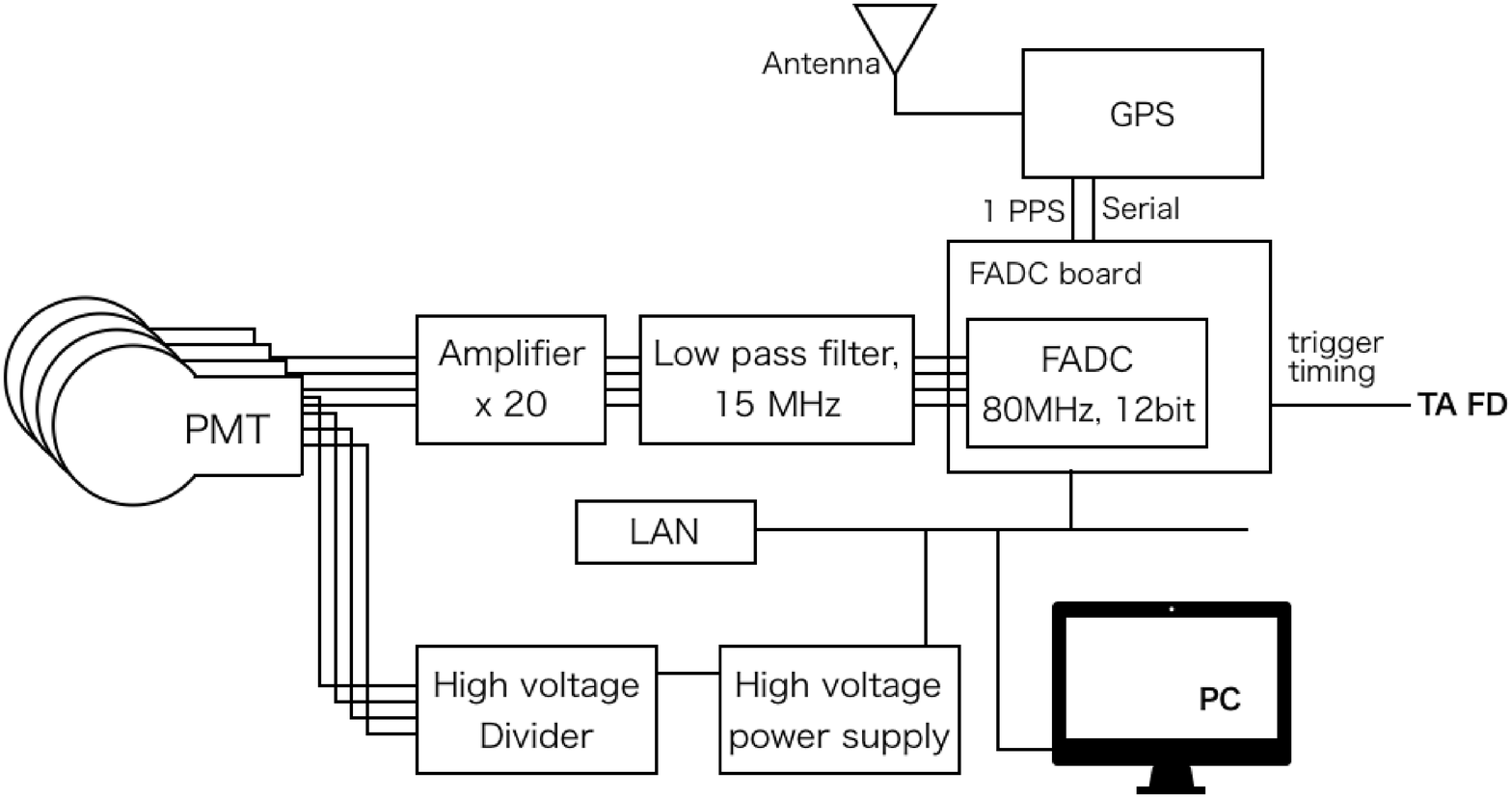}
  \caption{Block diagram of electronics and DAQ system of CRAFFT.}
  \label{fig-diagram}
\end{figure}

\begin{figure}[htbp]
  \centering
  \includegraphics[width=0.45\textwidth]{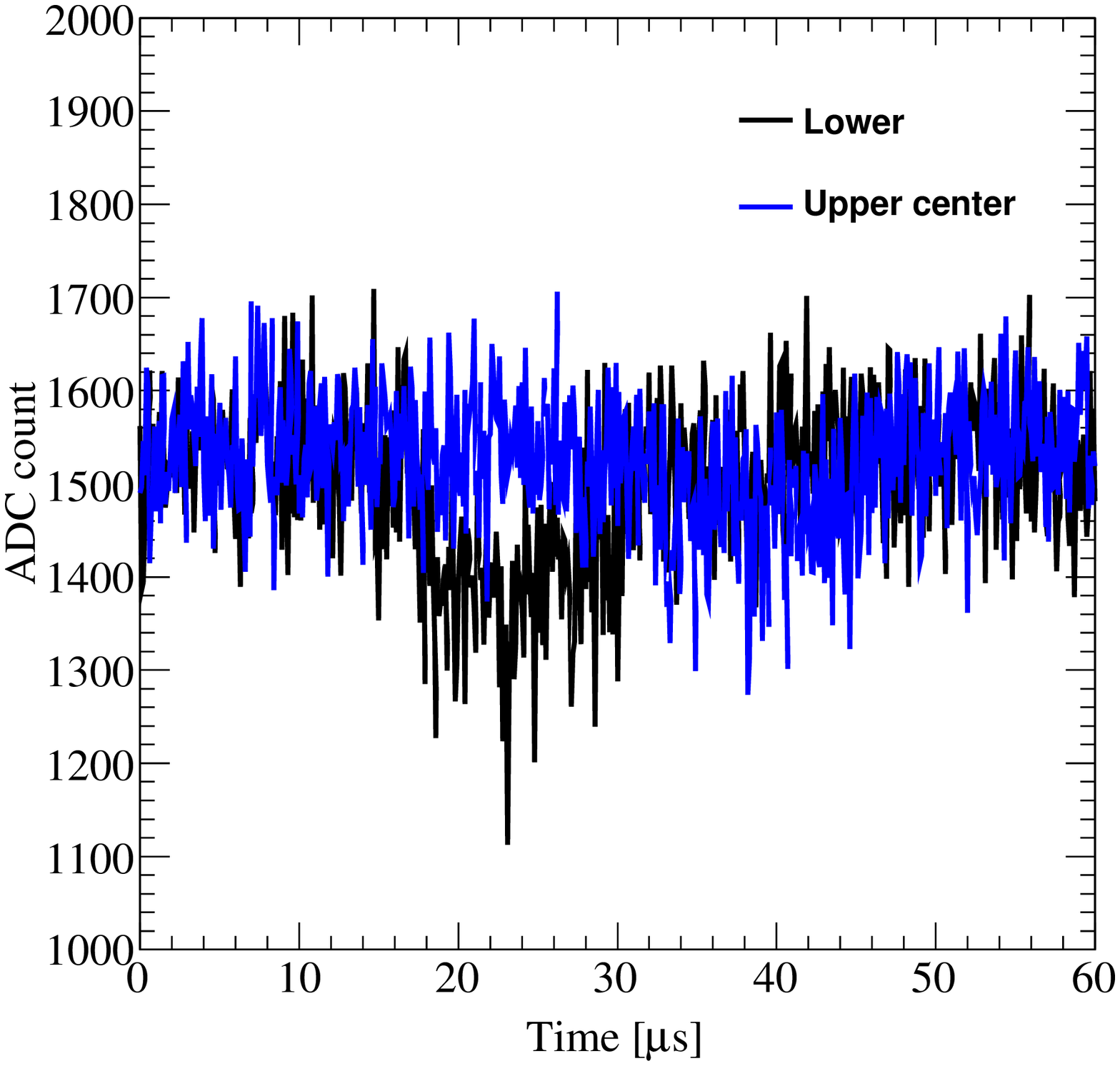}
  \includegraphics[width=0.45\textwidth]{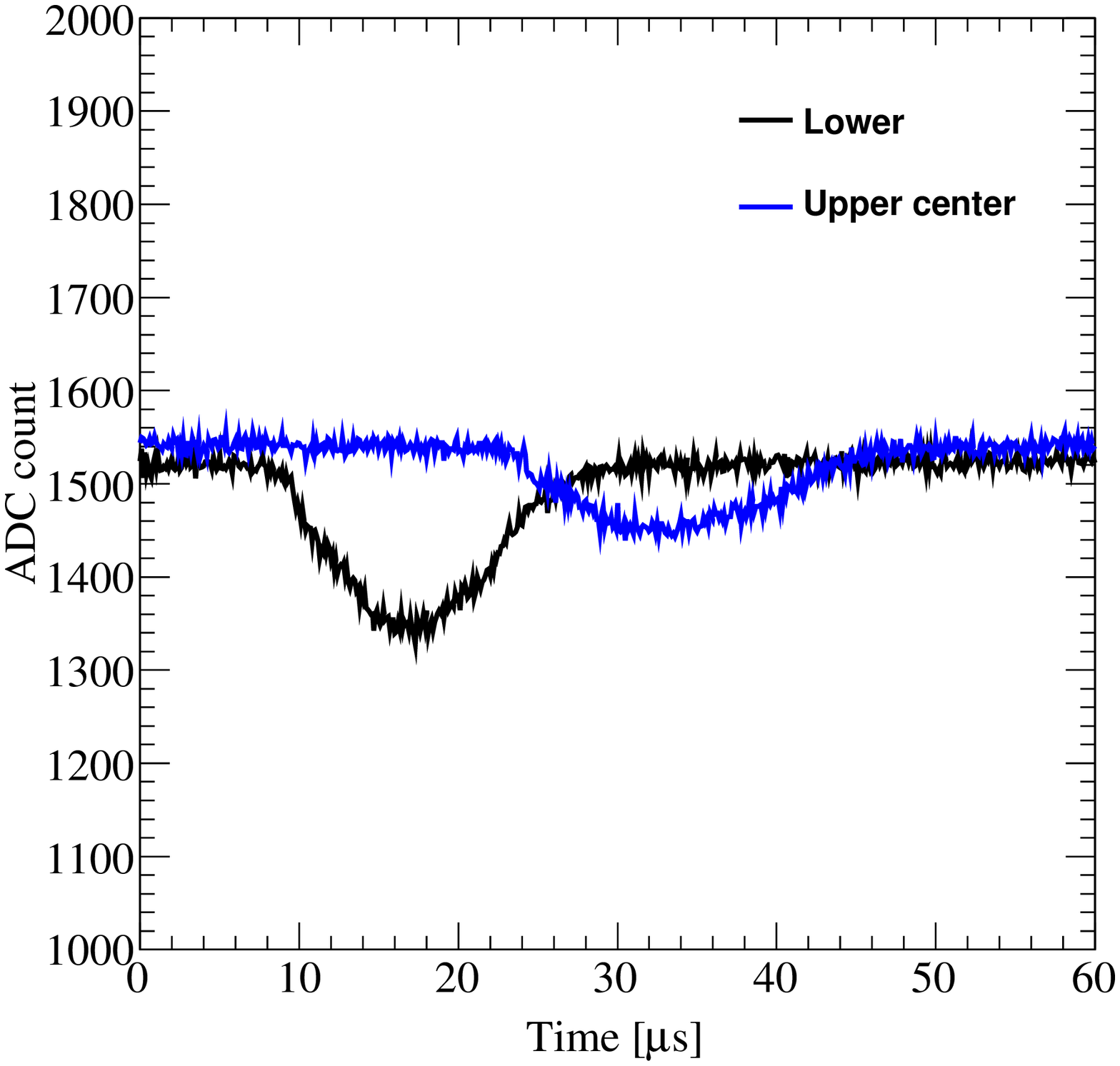}
  \caption{Waveforms of CLF detected by CRAFFT. Left: Typical
    single CLF event. Right: Averaged waveform of 133 CLF events.}
  \label{fig-clf}
\end{figure}

In 2017,
we conducted a test observation for ten days from November 9 to November 23,
when the TA FD was operating.
The total observation time was 63.4 h.

We operated CRAFFT remotely from the control room of the TA FD at the Black Rock Mesa site.
Figure \ref{fig-diagram} shows the block diagram of the electronics
and DAQ system of CRAFFT.
TA FDs recognize air shower events every $12.8~{\rm \mu s}$ and generate trigger timing pulses
that are distributed from the front panel of trigger electronics \cite{tameda}.
We used a trigger timing signal from TA FD for data collection.
We measured the relative gain of the telescopes,
including the transparency effect,
for both the lens and the UV filter,
using a UV LED mounted on the lens surface.
We regulate the applied voltage of PMT to adjust gain while measuring the LED light.
As a result,
we adjusted the gain of each telescopes within 20\%.
In this observation, we acquired 456,727 events. %%556,255
We searched air shower candidate events from all 456,727 data points recorded by CRAFFT. 
Most of the data did not include a signal 
because CRAFFT recorded the data using the trigger pulse from TA FD, 
the F.O.V. of which is about ten times larger than that of the CRAFFT detector. 
First, we selected the data with significant signals against the background. 
In addition, we excluded the CLF events that could easily identified from the time stamps.
Figure \ref{fig-clf} shows an example of a CLF event. 
Moreover,
air plane events were also removed using a pulse width that is much wider than that of typical air shower events.
Next, we selected the events with significant signals registered by more than two CRAFFT detectors. 
After this selection process, we obtained 6,600 event candidates.
Finally, we extracted signals like air shower event via eye scanning while considering pulse width and height.
We found ten apparent air shower events as a result of having compared those signals with corresponding TA Signals.

\begin{table}[htb]
  \centering
  \caption{The number of events that applied the selection.}
  \label{selection}
  \begin{tabular}{l|l}
    Number of acquired data &  456,727 \\ \hline
    Number of event candidate & 6,600  \\ \hline
    Number of clear air shower event &  10 \\
  \end{tabular}
\end{table}

\begin{figure}[htbp]
  \centering
  \begin{minipage}{1.0\hsize}
    \begin{center}
      \includegraphics[width=0.45\textwidth]{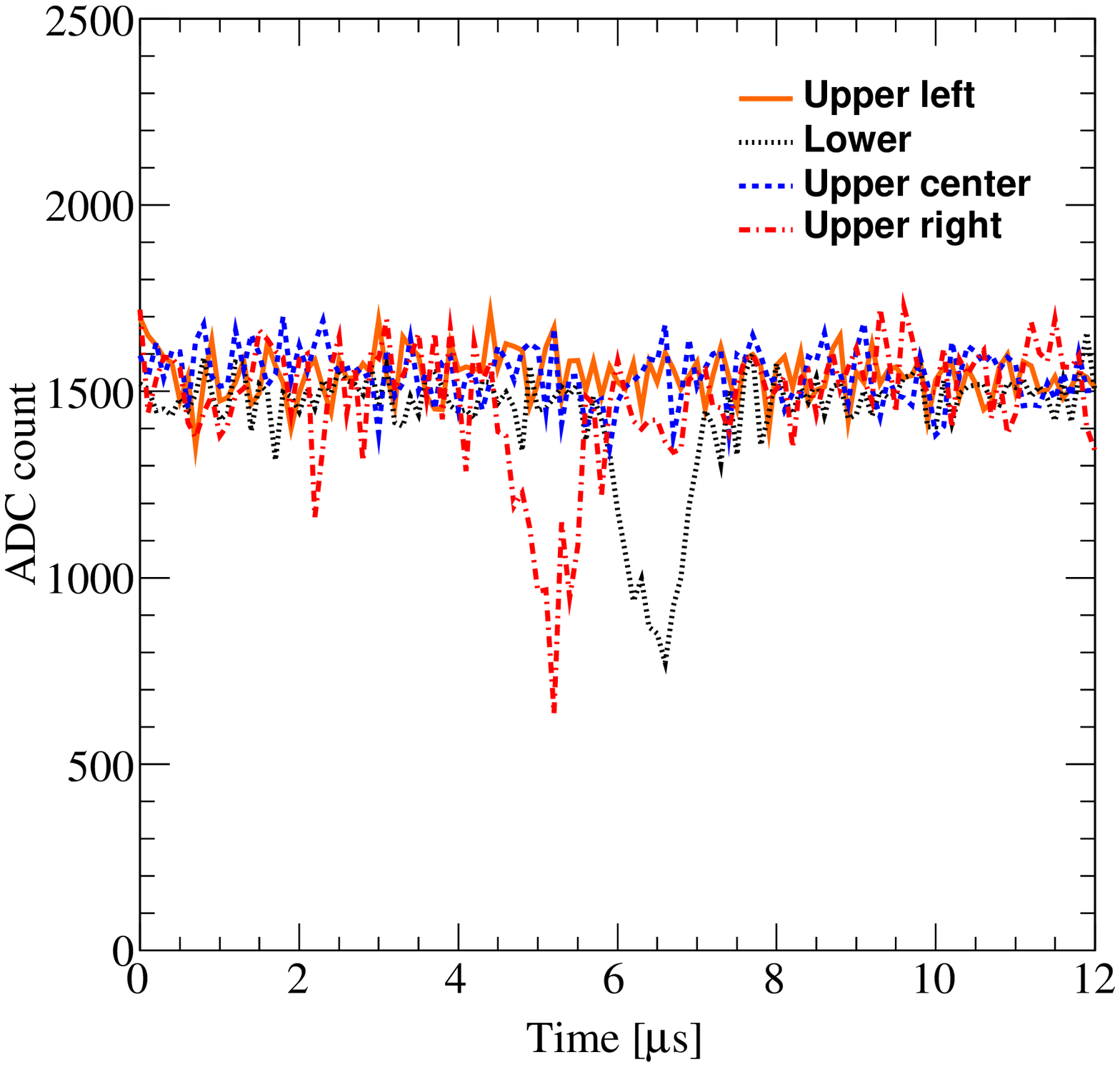}
      \includegraphics[width=0.45\textwidth]{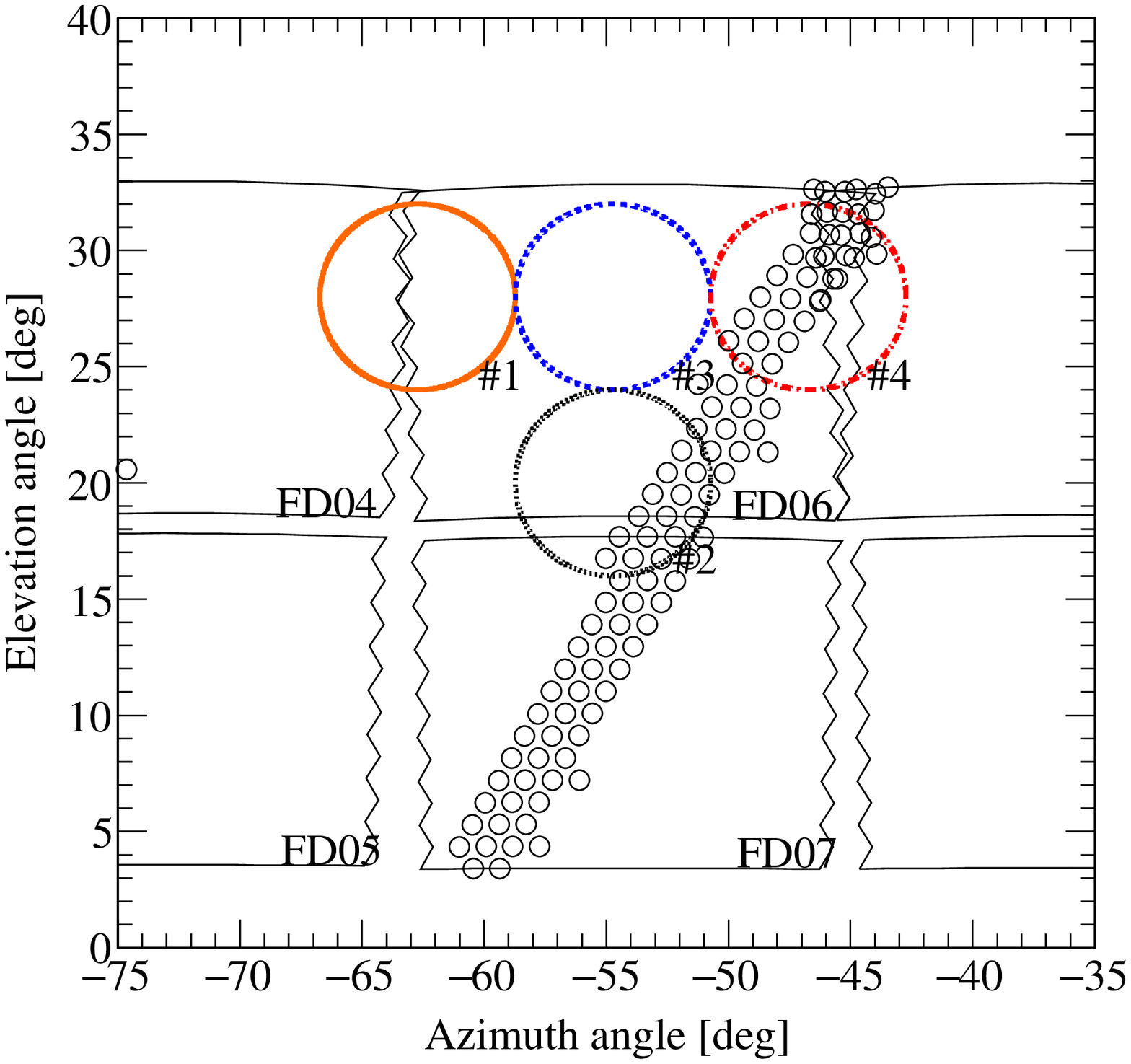}
      \hspace{1.6cm} [a] 2017/11/11 05:59:54.835620750
    \end{center}
  \end{minipage}
  \begin{minipage}{1.0\hsize}
    \begin{center}
      \includegraphics[width=0.45\textwidth]{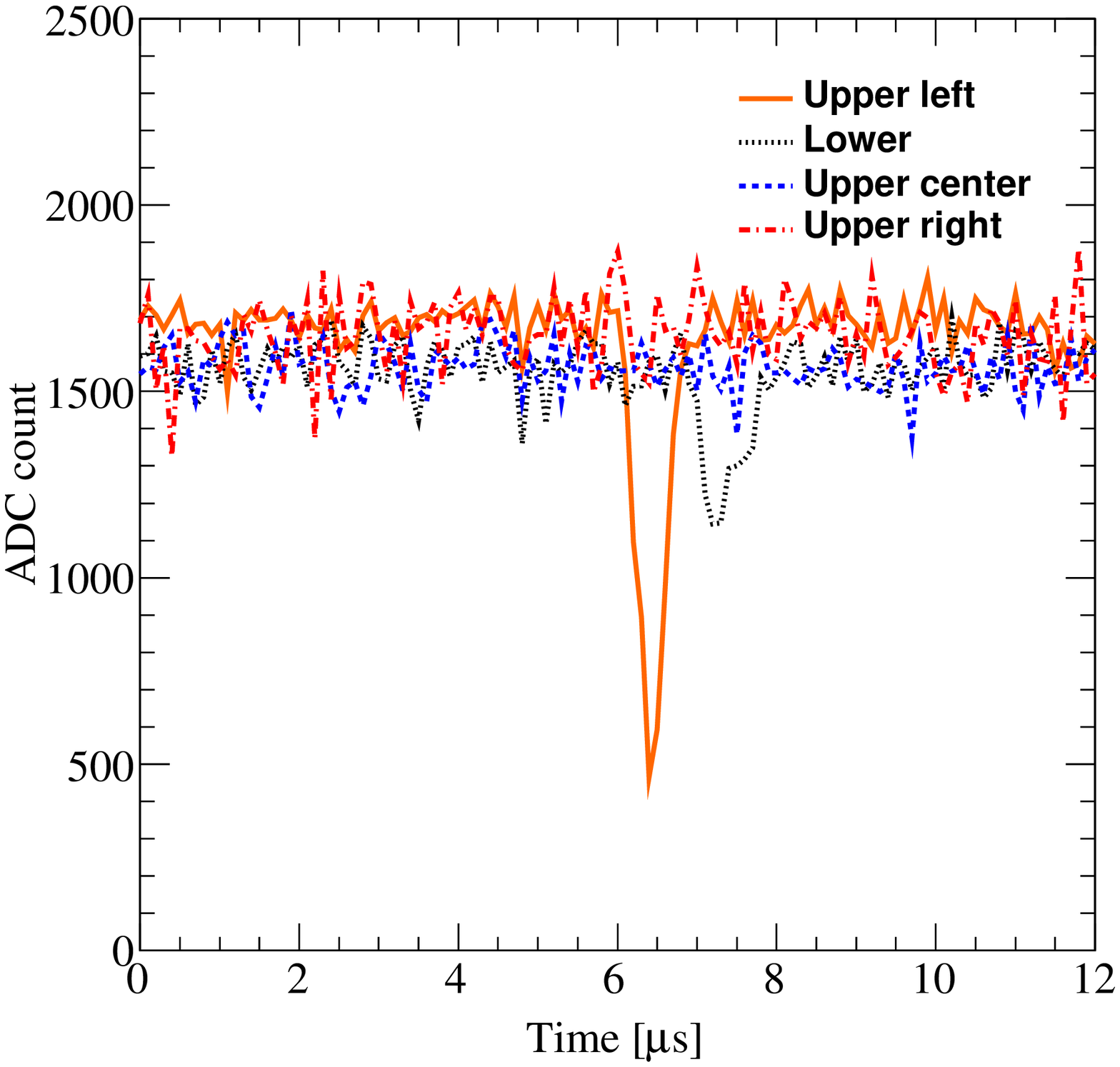}
      \includegraphics[width=0.45\textwidth]{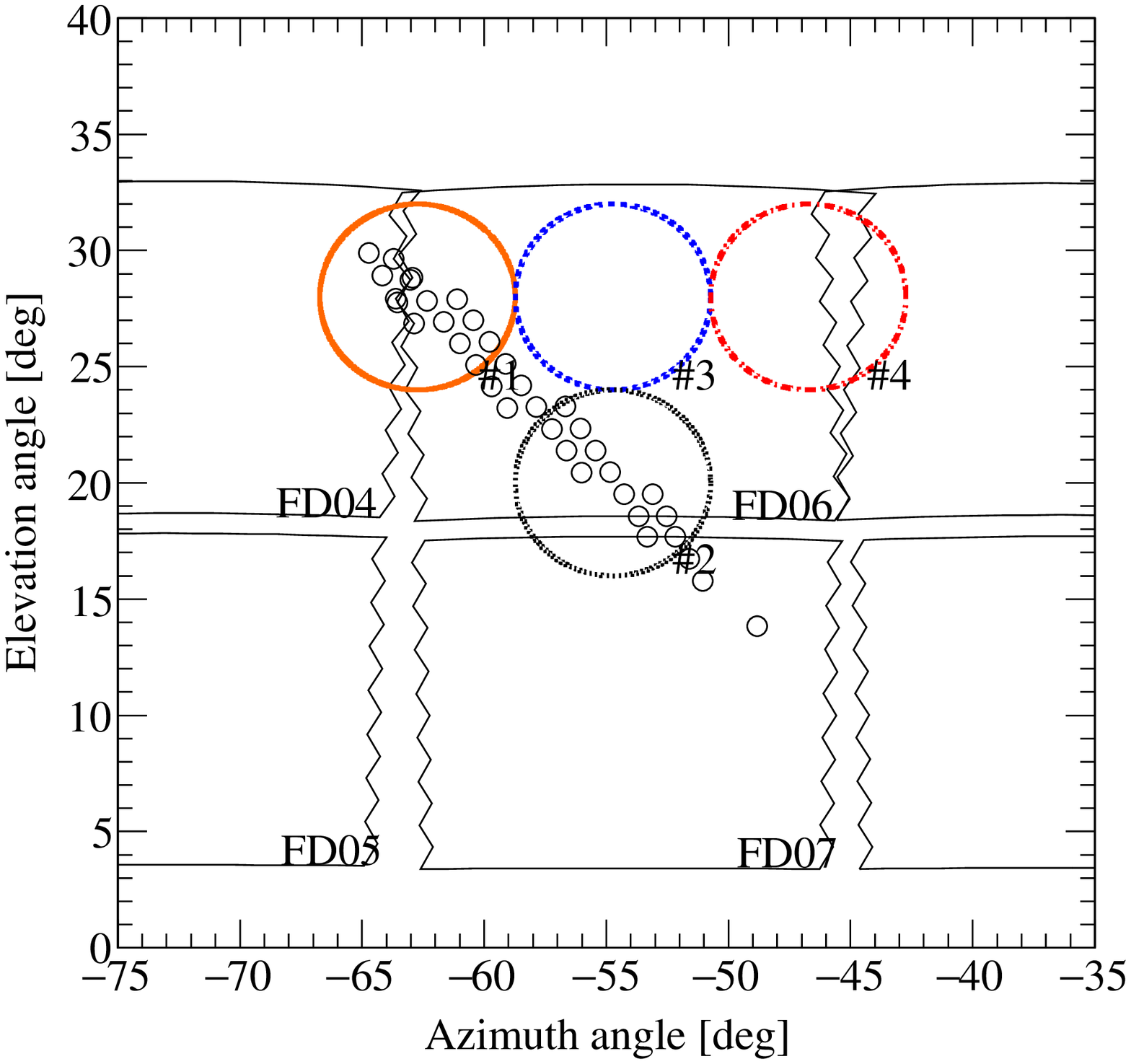}
      \hspace{1.6cm} [b] 2017/11/15 06:16:57.922469950
    \end{center}
  \end{minipage}

  \caption{Typical air shower event acquired by CRAFFT.
    Left: The FADC count of each channel 
    that can be compared within the accuracy of 20\%.
    Each bin of waveform is summed as $100~{\rm ns}$.
    Right: TA FD event display.
    Solid line is the edge of TA FD F.O.V. $1^{\circ}$ circles are views of TA FD triggered channels.
    $8^{\circ}$ circles are the F.O.V. of CRAFFT.
  }
\label{fig-events}
\end{figure}

Figure \ref{fig-events} shows some typical events that we successfully observed.
The left panels of Fig. \ref{fig-events} show waveforms of an air shower fluorescence signal recorded by CRAFFT.
The right panels show the corresponding event displays of TA FD.
As shown in Fig. \ref{fig-events},
we successfully observed air shower events that were also identified as air shower events by the TA FD. 

\section{Discussion}

We are developing a simple Fresnel lens FD, named CRAFFT, for UHECR observation.
We deployed four CRAFFT detectors at the TA site and performed test observations.
We acquired ten air shower events.
We successfully demonstrated that UHECR can be observed using CRAFFT,
which has a simple structure and an economic detector comprising only commercial products.
This can be an attractive option for application in next generation UHECR observatories.
In the case of a reflecting telescope especially for air shower observation,
a large phototube cluster can be an obstacle to screen incident lights.
On the other hand,
all of incident lights can be focused at the focal plane with a refracting telescope.
Therefore,
a refracting telescope can be smaller than a reflecting telescope to achieve the same light collection efficiency.
Additionally,
it is much easier to extend the field of view by enlargement of the focal plane.
The structure of a refracting telescope is very simple and compact
so that all the component can be covered easily to keep detectors clean.
This is good for easy maintenance. 
From these points, it is expected to impart the advantage of lower cost than ever.

FDs previously used for UHECR observation are expensive due to their strong structure
required to support a large composite mirror system and the multi-channel DAQ system.
Additionally,
detectors must be covered by a building.
In contrast,
a refracting telescope using a Fresnel lens with a single-pixel phototube,
such as the CRAFFT detector,
possesses a simple structure.
Thus,
the cost reduces by more than ten times than that of the previous detectors.
Moreover,
the CRAFFT detector is easy to deploy and can be installed without a hut
as all of its components can be covered by the structure alone due to its compactness.

We successfully demonstrated that a single-pixel Fresnel lens FD can be used for UHECR detection.
For future work,
we plan for stereo or multiple observations to establish a method to reconstruct geometry using CRAFFT.
We believe that the current configuration of CRAFFT can be further optimized.
We will also attempt the use of multiple pixels to improve $S/N$  or double lenses to extend the F.O.V.,
and reduce the cost per view.
CRAFFT is a promising detector that may significantly contribute
to the realization of next-generation UHECR observatories,
post TA$\times$4 \cite{tax4} or AugerPrime \cite{augerprime}.

\section*{Acknowledgment}

This work was supported by the JSPS KAKENHI Grant Numbers 25610051 and JP16K17710.
This work was partially carried out by the joint research program of
the Institute for Cosmic Ray Research (ICRR), The University of Tokyo.
This study was also supported by the Earthquake Research Institute The
University of Tokyo Joint Usage/Research Program.
The Telescope Array Collaboration supported CRAFFT as an associated
experiment and allowed us to use TA equipments
and FD event displays.
We wish to thank the staffs at the University of Utah, especially Prof. J.N.~Matthews.


%\bibliographystyle{ptephy}
%\bibliography{sample}
%
% once the .bbl file has been generated then place the text in your article.

\end{document}